# Microfluidic-Assisted Growth of Perovskite Microwires for Room-Temperature All-Optical Switching Based on Total Internal Reflection

Annalisa Coriolano[1†], Antonio Fieramosca[1†*], Laura Polimeno[1], Rosanna Mastria[1*], Francesco Todisco[1], Milena De Giorgi[1], Luisa De Marco[1], Aurora Rizzo,[1] Dario Ballarini[1], Ilenia Viola[1,2], Daniele Sanvitto[1]

[1]CNR NANOTEC, Institute of Nanotechnology, via Monteroni, 73100, Lecce, Italy

[2]CNR NANOTEC, Institute of Nanotechnology, c/o Dipt. di Fisica, Sapienza Università, P.le A. Moro 2, I-00185, Rome, Italy

**Abstract**

The development of efficient integrated photonic circuits is fundamental for ongoing research in information processing and computer science. The greatest challenge in strong light-matter regime facing photonic systems is achieving strong nonlinearities, which are exploitable in strongly coupled systems, leading to the formation of exciton-polaritons. In this context, the use of hybrid organic-inorganic perovskites offers a promising alternative, exhibiting robust interactions at Room Temperature (RT). However, the development of perovskite-based integrated devices requires both the ability to achieve long in-plane propagation, and the development of alternative fabrication approaches tailored to perovskite materials, designed to preserve their optical properties and prevent degradation.

Herein, we present the realization of a proof-of-concept all-optical switch using propagating polaritons confined in Total Internal Reflection (TIR), which ensures long in-plane propagation and limited optical losses. We realized an efficient injection/extraction of the TIR-confined waveguide polariton modes by employing gold grating couplers prepatterned on the substrate.

† These authors contributed equally to this work: A. Coriolano, A. Fieramosca.
* e-mail: antonio.fieramosca@nanotec.cnr.it; rosanna.mastria@nanotec.cnr.it;

## Introduction

Exciton-polaritons, hybrid light-matter bosonic quasi-particles resulting from the strong coupling between excitons and photons, exhibit distinct characteristics inherited from their individual constituents[1–3]. The photonic component contributes to a light effective mass (approximately $10^{-5}$ that of electrons), while the excitonic component introduces strong interparticle interactions (approximately $10^3$ times higher compared to standard nonlinear optical media)[4]. These peculiarities make them extremely sensitive to small changes in power, generally enabling long in-plane propagation, high switching efficiency, and no heat dissipation. Consequently, they fulfill the fundamental requirements for implementing all-optical computational schemes[5–9]. From a more strictly technologically relevant perspective, exciton-polaritons can also be easily formed by employing high binding energy materials[10] such as organics[11–13], transition metal dichalcogenides[14–16], and lead-halide perovskites[17–20].

In this regard, lead-halide perovskites (PVKs) have attracted significant attention due to their strong exciton binding energy, high luminescence quantum yield, and narrow emission linewidth. More interestingly, these unique features can be tailored by modifying the crystals composition and shape through precise control of synthetic conditions[21–23].

A plethora of fascinating results, such as polariton condensation, parametric scattering, and superfluidity, have been observed recently[24–28]. It has also been demonstrated that polariton interactions in perovskites at RT can be as high as those in GaAs-based microcavities at cryogenic temperature, making perovskites highly attractive for polaritonic applications[19]. The majority of these results have been obtained using a conventional planar microcavity, with two Distributed Bragg Reflectors (DBRs) sandwiching the active layer. However, state-of-the-art perovskite strongly coupled microcavities have high dissipation rates and short lifetimes, as well as low group velocity, thus limiting the maximum achievable in-plane propagation and preventing applications in cascaded-on-chip technologies. Therefore, there is a high demand for the use of different optical platforms to fully harness the potential of perovskite-based polariton. In this regard, perovskite single crystals, compared to their polycrystalline counterparts, exhibit superior properties owing to their reduced defect density, which enhances exciton lifetime and diffusion lengths[29], making them an ideal platform for photonic application. Alternatively to using planar microcavity and DBRs, perovskite single-crystals can support the formation of exciton-polaritons without the need for external resonators[18,30,31], due to the high refractive index contrast between them and the external environment. They can efficiently support propagating modes, such as waveguide modes, where optical confinement is achieved through Total Internal Reflection (TIR), making them highly suitable for the realization of active polaritonic circuits. The primary benefit of this system lies in the ability to achieve significant in-plane wavevectors, resulting in high group velocities, greatly exceeding those of planar microcavities with an equivalent lifetime. Moreover, this configuration requires less time for fabrication, minimizes the mode volume, and reduces optical losses.

The significant interest in this direction is evidenced by the recent development of perovskite-based waveguide couplers, interferometers, and beam splitters[18,32–34]. To achieve this, the shape and quality of the crystals must be appropriately tuned to obtain sharp edges, uniform surfaces, and geometric parameters that support TIR and minimize optical losses. A key factor in obtaining perovskite single crystals with desired shape and features is the confinement of the precursor solution. In recent years, various synthetic approaches have been developed to tailor the dimensions and shapes of perovskite single-crystals on arbitrary substrates[18,31,35–38]. Among these, microfluidic assisted confined crystal growth has emerged as a highly promising and effective method for producing high-quality perovskite single-crystal of desired shape[17,18,39].

However, although perovskite based optical circuitry represents a highly promising photonic platform, the crystals developed until now lack efficient injection/extraction, since the TIR-propagating signal is collected from the edges. A robust design necessitates efficient light coupling achieved through a grating coupler, which typically involve post-processing steps utilizing polar solvents that may impact the quality of the perovskite[40,41].

In this work, we overcome these limitations by leveraging the microfluidic-assisted growth technique to directly fabricate a 2D perovskite waveguide on a gold grating. This technique enables precise control over the dimensions, shape, and position of the active material on the final device, thereby removing constraints associated with fabricating an optical grating using post-processing fabrication techniques[42,43].

The effectiveness of this method is demonstrated by the synthesis of high-quality dodecylammonium lead halide PVKs microwire (MW), specifically $(C_{12})_2PbI_4$ ($C_{12}$ = dodecylammonium, $C_{12}H_{25}NH_3^+$), consisting of alternating layers of $PbI_6^{4-}$ octahedra layer interspaced with a double layer of C12. The long alkyl chain organic spacer enhances the environmental stability of the crystal while also ensuring a high exciton binding energy[44–46]. Employing a single grating configuration, we demonstrate how waveguide polaritons, off-resonantly excited, can propagate over considerable distances into the perovskite microwire and be efficiently extracted by the gold grating placed far from the excitation spot. Moreover, by exploiting the exciton-polariton nonlinearities and a coherent resonant excitation in a double grating configuration, we realize a proof-of-concept polariton switch operating in TIR. Our approach demonstrates a fully integrable, robust, and reproducible polariton device, showcasing significant potential for practical applications in advanced photonic circuits and optoelectronic systems.

## Results and discussion

The sample structure consists of C12 perovskite microwires grown on a substrate with gold gratings fabricated on commercial glass. The chemical structure of $C_{12}$-perovskite is depicted in Figure 1a. The inorganic layers of $PbI_6$ octahedra are intercalated by long alkyl chains, resulting in a separation of 24.437 Å between the two nearest least squares planes passing through the equatorial atoms of the $PbI_6$ octahedra [47]. The presence of a long ammonium chain can improve device stability under environmental conditions[48]. Initially, a layer of polymethylmethacrylate (PMMA) is initially spin-coated onto the glass substrate, followed by electron-beam lithography of the grating and subsequent development. Then, 60 nm of gold is thermally evaporated, followed by a lift-off process in acetone (see Supporting Information for further details). Using a homebuilt micromanipulator (Figure S1), a patterned polydimethylsiloxane (PDMS) replica with microchannels having a height of 500 nm and variable widths is aligned with the gold gratings (Figure 1b), bringing the substrate in conformal contact with the PDMS replica. $C_{12}$-perovskite microwires are prepared using a microfluidic-assisted technique from a precursor solution, which is deposited at one end of the microchannels (Figure 1c). Capillary forces guide the solution into the microchannels, filling them and ensuring the formation of $C_{12}$ microwires through confined nucleation. The growth process is fine-tuned by optimizing key parameters such as temperature, precursor solution concentration, solvent type, and ambient solvent saturation. This optimization is essential for obtaining sharp edges and low roughness, reducing scattering losses, and enabling long-range polariton propagation. To achieve high crystal quality, a 0.25 M stoichiometric solution of precursors in gamma-butyrolactone has been used, with the growth process conducted at RT. During the nucleation phase, the substrate with the template filled with the precursor solution is placed in a box sealed with parafilm. This setup creates a solvent-saturated environment that slows the nucleation phase, promoting the formation of a low number of nuclei. These nuclei evolve into perovskite microwires as the solvent of the precursor solution gradually evaporates (Figure 1d). After 8 hours, the growth is complete, and the PDMS replica is removed. Millimeter-long perovskite microwires are formed, precisely positioned on the substrate atop of the gold gratings, as shown in Figure 1e. Scanning electron microscope images of the as-synthesized microwires highlight the high quality of C12 crystals, which feature a well-defined shape determined by the PDMS template (Figure S2).

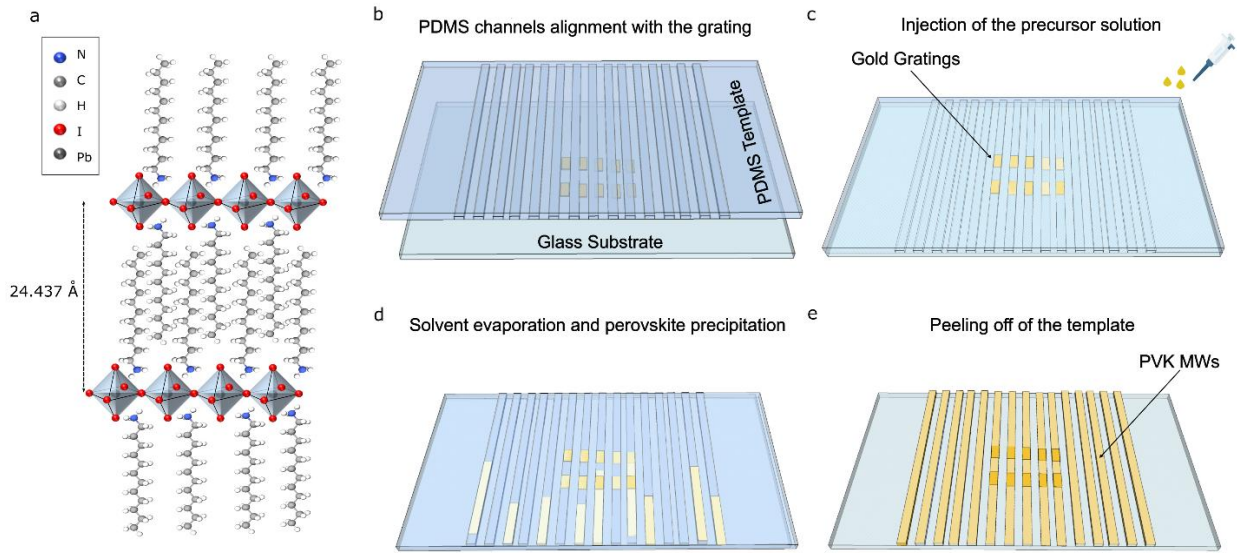

**Figure 1.** (a) 3D schematic of the C12 perovskite structure. Illustration of the template-confined growth process for C12 microwires. First, the patterned PDMS template is aligned with the gold gratings (b) and transferred onto the glass substrate. The precursor solution is then injected into the PDMS microchannels (c), initiating solvent evaporation and perovskite precipitation (d). After 8 hours, the solvent is fully evaporated, resulting in the formation of perovskite microwires. Once the template is peeled off, single crystal microwires remain on the glass substrate (e).

The periodicity of the grating is designed to allow for the extraction of the signal confined in TIR within the perovskite microwires. Rigorous coupled wave analysis (open-source S4 software[49]) is employed to simulate the optical response of the grating, considering a $C_{12}$-perovskite microwire with a thickness of 500 nm grown on top. The real and imaginary parts of the perovskite refractive index are taken from ref.[50]. The theoretical energy versus in-plane momentum ($k_{//}$) reflectivity map, considering Transverse Electric (TE) polarized light, is reported in Figure 2a. The formation of strongly coupled propagating and counter-propagating lower polariton (LP) waveguide modes is clearly visible. The characteristic dispersion of the waveguide modes, which is linear in the weak coupling regime, exhibits a notable increase in curvature as the energy nears the exciton resonance ($E_{exc}$ = 2525 meV, indicated by the white dashed line in Figure S3), thus providing clear evidence of the achievement of strong coupling.

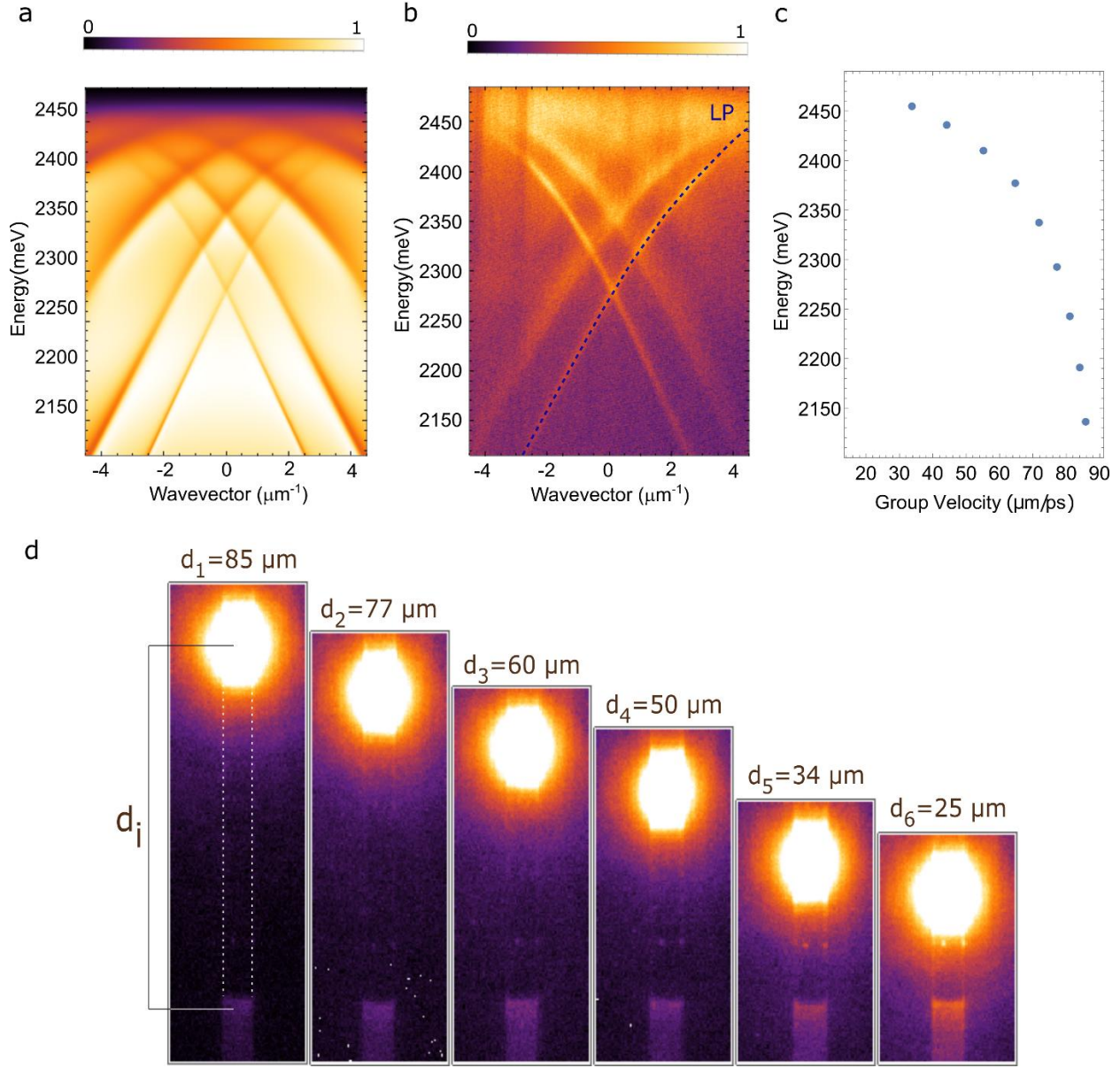

**Figure 2.** (a) Theoretical energy versus in-plane momentum reflectivity map, considering TE polarized light. (b) Experimental energy versus in-plane momentum reflectivity map, considering TE polarized emitted light. The extracted Rabi Splitting is 340 meV. The dashed blue line indicates the dispersion of one of the manifold LP waveguide modes. (c) Group velocity as a function of energy for the mode marked by the dark blue dashed line. (d) Real-space PL maps as a function of the distance $d_i$ between the excitation spot and the outcoupler grating.

The formation of the upper polariton (UP) branch can be seen in Figure S3, where the simulation is performed in a broader energy region. The appearance of multiple LP waveguide modes is a direct consequence of the free spectral range which, with 500 nm of perovskite, supports the formation of multiple waveguide modes. Moreover, the dispersion of the bare grating in the same spectral region and TE polarization is shown in Figure S4 and does not display the formation of any lattice mode, thus confirming that the adopted design effectively works as an out-coupler for the TIR-confined waveguide polariton modes. Experimentally, the grating is directly tested using momentum resolved photoluminescence (PL) measurements. The grating region is excited in the center by using a continuous wave (CW) off-resonant pumping (λ=405 nm) and the PL signal collected in TE polarization. The resulting energy versus in-plane momentum ($k_{//}$) PL map is shown in Figure 2b and it is in good agreement with the theoretical calculation. It is important to note that the visibility of the UPB is hindered by the strong absorption at high energy, as previously reported[30,47,51,52]. From the fit of the experimental data (blue dashed line in Figure 2b), we obtained a Rabi splitting of $\Omega_{TE}$ = 340 meV, which is in good agreement with previous reports[18,53,54]. Moreover, with reference to the mode highlighted in Figure 2b, we calculated the group velocity of the mode as a function of energy, $v_g(E)$, evaluated as follows: $v_g = 1/\hbar * \partial E/\partial k$. Depending on the excitonic fraction, the group velocity ranges from 30 to 90 μm/ps, that is one order of magnitude higher than standard microcavities[7] and comparable to propagating surface modes, such as Bloch Surface Waves (BSW), employed in the past to strongly couple a $WS_2$ monolayer and organic materials[55,56].

As discussed above, waveguide modes are capable of harnessing propagation properties. However, it is crucial for perovskite microwires to exhibit minimal scattering to preserve the benefits of working with TIR-confined optical modes. To assess the quality of the perovskite microwires, we employed a single grating configuration wherein the material is off-resonantly excited at various distances ($d_i$) from the out-coupler grating, as reported in Figure 2d. From the real space PL maps, we have estimated the emission intensity collected below the outcoupling region for each distance, $I_{d_i}^{Out}$. With this approach we evaluate the propagation loss coefficient α, following the relation: α = $(1/L)$ 10 log ($I_{d_i}^{Out}/I_{d_0}^{Out}$), where $I_{d_i}^{Out}$ is the intensity at every distance, $I_{d_0}^{Out}$ is the intensity at the shortest distance, and $\Delta L = (d_i - d_0)$ the relative distance, with $d_0$ taken as the reference point for the shortest distance. Therefore, we have extracted a loss coefficient α for each pair of measurements, as reported in Figure S5 and obtained an average value of α = 0.12 +/- 0.03 dB/um. This low value is a direct consequence of the microfluidic-assisted growth method employed here, which significantly reduces impurities and ensures sharp edges and long-range homogeneity. Compared to similar wires, such as $CsPbBr_3$ wires grown via the capillary bridge method and integrated into planar microcavities[7], it appears slightly lower, thereby confirming how the additional deposition steps affect the optical losses.

Although the PL measurements discussed above clearly confirm the excellent optical quality of the perovskite microwires, the realization of on-chip integrated devices and circuits relies on the efficient injection and extraction of a coherent propagating signal. Recently, various optical structures such as waveguide couplers, interferometers, beam splitters, X-couplers, Y-couplers, have been theoretically investigated and experimentally realized in perovskite films using techniques such as Focused Ion Beam lithography or assembling perovskite quantum dots with a capillary bridge technique[32,33,57]. In these studies, coherent propagation is achieved through optical pumping of the active material above the lasing threshold. However, these implementations lack efficient injection/extraction, as the structures are pumped off-resonantly, and the TIR-propagating signal is collected from the edges. Here, we implement an ideal scheme in which a resonant injection and the extraction of the TIR-propagating signal is performed with two gratings.

The perovskite microwires are grown on top of two identical gratings, spaced approximately 10 μm from each other, as depicted in the optical image presented in Figure 3a. The top grating is utilized to perform a resonant injection of a CW laser with a photon energy of 2330 meV and a well-defined in-plane momentum, $k_{//} = 1.12$ μm$^{-1}$, resulting in a group velocity of $v_g = 62.27$ μm ps$^{-1}$. The real space map is shown in Figure 3b, while the pumped area is indicated by the green circle in the zoomed-in energy versus $k_{//}$ map displayed in Figure S6. By carefully matching the in-plane momentum of the waveguide mode, it is possible to obtain a propagating polariton flow with a well-defined velocity that is outcoupled from the second grating.

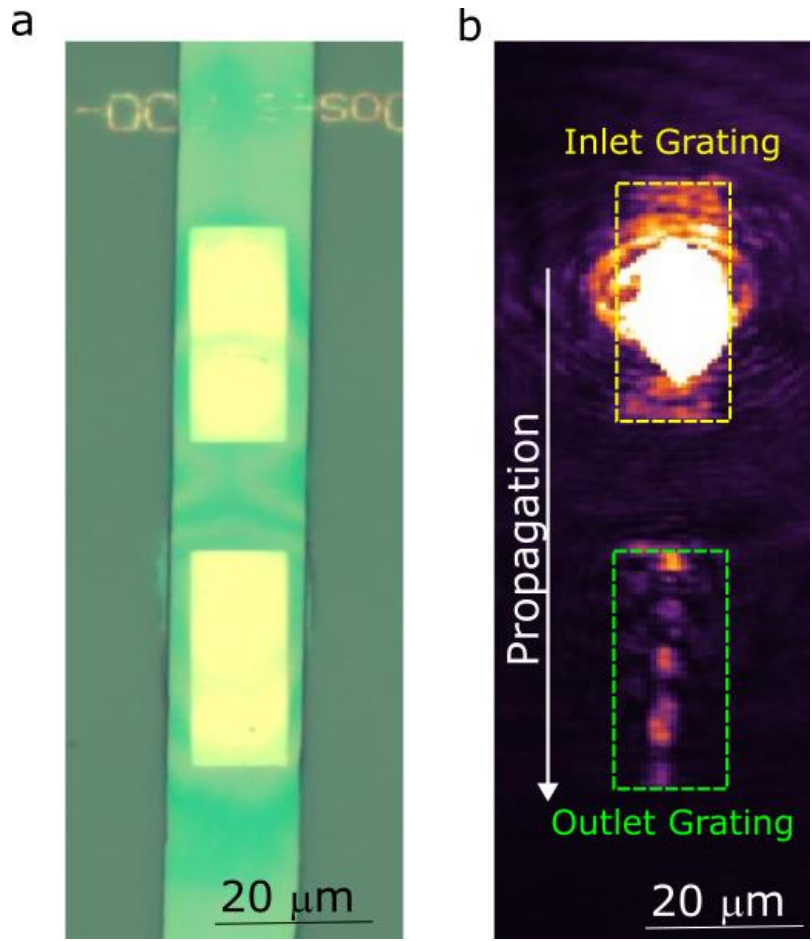

**Figure 3.** (a) Optical image of C12 microwire grown on the top of gold gratings. (b) Real-space resonant propagation. The excitation spot is placed on the top intel inlet grating (dashed yellow rectangle) while the bottom grating (dashed green rectangle) outcouples the radiation. The white arrow indicates the propagation direction.

It is important to note that the outcoupled signal shows an interference pattern on the grating region. This is a direct consequence of the laser coherence, which, once it reaches the out-coupler region, backscatters at the opposite in-plane momentum, thereby creating the interference pattern visible in real space. Note that such interference pattern demonstrates the long propagation even under the diffraction grating.

This is in contrast with the real space PL maps shown in Figure 2d, which, instead, display a uniform signal which spreads over the whole grating region.

Based on the results shown above, we experimentally demonstrate that our samples exhibit a robust design characterized by long in-plane propagation and high velocities. Consequently, it is natural to explore experimental configurations capable of realizing logic operations, thereby harnessing polariton nonlinearities. Within the exciton-polariton framework, various proof-of-concept logic elements have been successfully implemented in GaAs-based planar microcavities[5,58]. However, cost-effective and technologically relevant polariton devices require RT operation, presenting a significant challenge for conventional inorganic semiconductors due to their low exciton binding energy. To date, only a handful of polariton devices operating at RT have been demonstrated. Organic microcavities and TMDs have been utilized to implement all-optical polariton transistor and switches, although without making use of in-plane propagation[11,59,60]. Conversely, the full utilization of propagation properties has only recently been realized through the development of a nonlinear source of polaritons based on propagating BSWs in strongly coupled $WS_2$ monolayer[55], as well as an optical switch based on planar microcavities incorporating perovskite microwires[7]. In this context, our platform offers an optimal trade-off between minimizing footprint and harnessing in-plane propagation. To this end, we initially investigated the existence of a nonlinear response under optical pumping in our samples (Figure S7). One of the two gratings is resonantly pumped using a pulsed laser (2250 meV, 10 kHz, 100 fs) in a reflection configuration. The laser is kept broad in momentum space, and the LP waveguide modes are discernible as dips in the laser's spectrum, as visible in the energy vs $k_{//}$ reflectivity map collected at low excitation power and reported in Figure S7b. By increasing the pumping power, a clear blueshift becomes evident, as shown in Figure S7c. This observation aligns well with previous findings demonstrating a pronounced nonlinear response in various strongly coupled perovskites[19,61–63]. It is important to note that the total energy shift exceeds the linewidth of the LP waveguide modes (See Figure S7d and e), therefore satisfying the necessary conditions for implementing a polariton-based all-optical switch. To fully exploit the proven strong non-linearities in a real device, we present an experimental demonstration of this proof-of-concept all-optical switch. For this specific experiment, we employed two lasers: a fs pulsed laser, which shifts the energy position of the LP waveguide modes, and a CW laser, acting as the propagating signal within the microwire. The CW laser ($E_{CW}$=2330 meV) is focused at a given in-plane momentum and slightly off-resonance with respect to the LP waveguide mode. The off-resonance condition is achieved by precisely tuning the CW laser to an in-plane momentum slightly smaller than that of the LP waveguide mode at the same energy ($k_{//} = 1.12$ μm$^{-1}$). In real space, the Full Width at Half Maximum (FWHM) of the CW laser spot is approximately 3 μm. The pulsed laser, with a linewidth of about 15 meV and centered around $E_{pulsed} = 2290$ meV, is on-resonance with the LP waveguide mode. To prevent any undesired propagation of residual signals from the pulsed laser at the out-coupler grating, it is configured to excite the LP waveguide mode propagating in the direction opposite to that of the CW laser. This experimental configuration allows us to exclusively monitor the modifications induced by the pulsed beam on the propagating CW laser. A scheme of the configuration is depicted in Figure 4a. Then, we evaluated the intensity emitted below the out-coupler grating (dark green square in Figure 4a) and compared it with the intensity immediately before (dark blue square in Figure 4a), i.e., the residual scattered signal not associated with the propagating CW beam. We performed this analysis as a function of the pumping power of the pulsed laser, as reported in Figure 4b for region 1 (dark green points) and region 2 (dark blue points), respectively. Under these experimental conditions and at low pumping power, the system is in the OFF-state regime. This means that the signal collected below the out-coupler grating, associated with the propagating CW laser, is negligible, as very little light is coupled into the propagating mode. This low-power regime is evidenced in Figure 4b by the red band on the left side of the figure. In contrast, as the pump power is increased, the LP waveguide modes undergo a blueshift, and the CW laser becomes resonant with the propagating mode. Consequently, polaritons are injected below the in-coupler grating, propagate, and are detected below the out-coupler grating, i.e., the system reaches the ON-state regime. This high-power regime is depicted in Figure 4b by the yellow band on the right side of the figure. The nonlinearity results in a clear "kink" in the signal detected below the grating, distinctly different from the scattered signal. This two-beams experiment effectively proves how the designed sample can act as an all-optical polariton switch and demonstrate the realization of an active polariton waveguide.

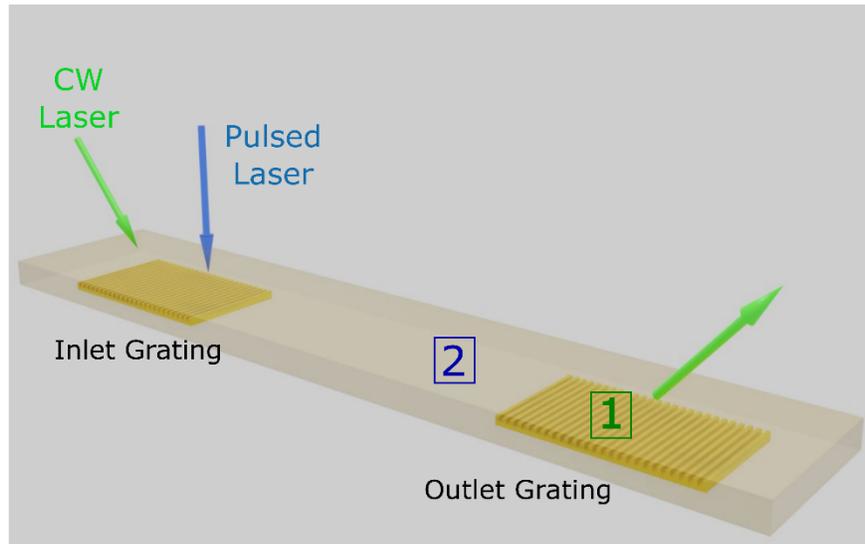

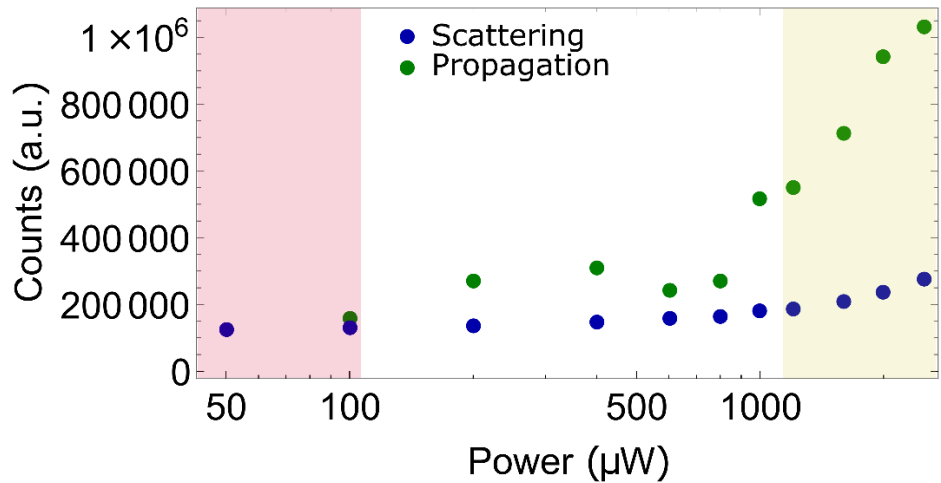

**Figure 4.** (a) Sketch illustrating the optical measurement configuration where a fs pulsed laser (blue arrow) allows the gating of a CW laser (green arrow), therefore acting as an ON-OFF switch of the propagating signal within the microwire. (b) Power-dependence of the intensity (counts) of the propagating signal collected in region "1" (dark green dots) and the scattering signal (dark blue dots) collected in region "2" of the microwire. The red and yellow bands indicate the OFF and ON state regime, respectively.

## Conclusions

In conclusion, we have successfully demonstrated the realization of an active polariton waveguide based on propagating polaritons self-sustained by a dodecylammonium lead halide perovskite microwire. A microfluidic-assisted growth technique is employed to obtain waveguide-based devices, showcasing the capability to effortlessly tune the properties of the final device by adjusting dimensions, shape, and position of the active material. Thanks to the capability of this technique, we achieve efficient injection and extraction of the TIR-propagating polaritons by directly growing a 2D perovskite microwire on a gold grating. We leverage the excellent propagation properties of waveguide polaritons along with the strong polariton-polariton interactions observed in these materials to successfully realize an all-optical switch operating at RT. Our approach demonstrates the realization of a robust and reproducible polariton device with minimal footprint, capable of full integration without the need for additional post-processing steps that may degrade the quality of the active material.

**Methods**

**Sample fabrication**

*Precursors solution*

n-dodecylammonium iodide, Acetone, Isopropanol (IPA), are purchased from Sigma Aldrich, Polydimethylsiloxane (PDMS) Sylgard 184 is purchased from Dow-Corning. Lead(II) iodide ($PbI_2$) is purchased from Alfa Aesar. γ-butyrolactone (GBL) is purchased from TCI. All chemicals are used as received without any further purification. The precursor solution is prepared within a $N_2$-filled glovebox by dissolving n-dodecylammonium iodide (156.63 mg, 0.5 mmol) and lead iodide (115.25 mg, 0.25 mmol) in 1 mL of γ-butyrolactone. This process yielded a 0.25 M solution, which is stirred at 70 °C for 1 hour until achieving a clear yellow solution.

*PDMS template*

A micro-structured PDMS template, featuring channels with dimensions h: 500 nm and w: 2-20 μm, is created by blending the prepolymer and curing agent in a 10:1 weight ratio. The mixture is then placed in a desiccator for 30 minutes to eliminate any bubbles. Subsequently, the liquid PDMS is cast onto a commercially patterned silicon master (ThunderNIL). After meticulous removal of air bubbles, the PDMS is cured in an oven at 140°C for 15 minutes. The elastomeric replica was then detached from the master and placed in conformal contact with the device substrate, with the microchannels open to facilitate the microfluidic injection of a small volume of precursor solution.

*Gratings fabrication*

Glass substrates are cleaned thoroughly with acetone and IPA in an ultrasonic bath, followed by a dehydration bake of 10 minutes at 180°C on a hotplate. Polymethyl methacrylate (PMMA) is used as an electron beam resist (MicroResist, 950 PMMA A4) and is spin-coated on cleaned samples at 3000 rpm for 60 seconds. Subsequently, the samples are annealed at 180°C for 180 seconds. Finally, in order to avoid surface charging during electron beam lithography, a conductive polymer is spin coated on the samples (MicroResist, DisCharge $H_2O$) at 2000 rpm for 60 seconds, followed by a soft baking at 90°C for 60 seconds.

E-beam lithography is performed on a 30 kV column system (Raith). Written samples are then immersed in water to remove the thin conductive polymer and then finally developed in a methyl isobutyl ketone:IPA 1:3 solution for 60 seconds. Finally, 3 nm of chromium and 60 nm of gold are thermally evaporated onto the substrates, followed by overnight lift-off in acetone at room temperature.

*Perovskite growth*

The alignment of the patterned template with the gold gratings on the glass substrate is achieved using a custom-built micromanipulator in coordination with the PDMS channels. To initiate the deposition process, a 1 µL drop of the precursor solution is placed at one end of the microchannels within the PDMS template. Capillary forces guide the solution, allowing it to fill the channels of the PDMS template seamlessly. The entire setup is then enclosed in a sealed box, secured with parafilm, and left at room temperature. As the solvent gradually evaporates, crystal precipitation occurs within the channels of the template. After 8 hours, the PDMS template is carefully removed, resulting in high-quality C12 microwires perfectly aligned with the gold gratings on the glass substrate.

**Optical Measurements**

All the optical measurements are performed in ambient conditions at RT. For the photoluminescence measurements, the microwire is off-resonant excited by using a CW diode laser ($\lambda$ = 405nm). The photoluminescence is collected in reflection configuration, using a long-working distance objective (Olympus, 40x/NA=0.6). The angle resolved dispersion is taken by imaging the back focal plane of the objective onto the entrance slits of a spectrometer (Princeton Instruments, Acton Spectra Pro SP-2300) equipped with three gratings (150 lines/mm, 300 lines/mm, 600 lines/mm) and coupled to a 2D charge-coupled device array (Princeton Instruments, Pixis 400).

For the all-optical switching measurements, two sources are employed: a tunable femtosecond laser and a CW diode laser operating at 532 nm. The pulsed source is derived from a Ti:sapphire laser (Coherent, Vitara) that is pumping an ultrafast amplifier (Coherent, Legend). The amplifier is coupled with a computer-controlled Optical Parametric Amplifier for wavelength modulation (Coherent, TOPAS). The repetition rate and pulse width are 10 kHz and 100 fs respectively.

It is important to note that the pulsed laser is responsible for opening temporal windows that allow the propagation of the CW laser at 10 kHz. Although the temporal width of these windows also depends on the polariton lifetime and the microscopic mechanisms driving the nonlinear response[64], the repetition rate is relatively low, and the switching effect can only be observed in time-integrated measurements, averaged over several pulses, in order to accumulate sufficient propagating counts on the output grating. For this reason, the data presented in Fig. 4 are integrated over 20 seconds for each pumping power. Additionally, the choice of the initial state is crucial. In our experiments, we always performed switching-ON measurements, meaning that at low pumping power, no counts are associated with the propagating signal on the output grating. This is advantageous for the measurements, as it ensures a clear distinction between the signal at low and high pumping powers.

**Simulation and fits**

The dispersion patterns presented in Figure 2a are modeled using the semi-analytical Rigorous Coupled-Wave Analysis (RCWA) method as implemented by the S4 package.

The experimental data depicted in Figure 2b are fitted utilizing a 2x2 coupled harmonic oscillator Hamiltonian:

$$\begin{pmatrix} E_{ph}(k) & \Omega/2 \\ \Omega/2 & E_{exc} \end{pmatrix}$$

Here, $E_{exc}$ represents the exciton energy, $\Omega$ denotes the Rabi splitting, and $E_{ph}(k)$ is the energy dispersion of the photonic branch.

**Acknowledgements**

The authors gratefully thank P. Cazzato and S. Carallo for technical support.

This work was financially supported by the PNRR MUR project 'National Quantum Science and Technology Institute' – NQSTI (PE0000023); the PNRR MUR project 'Integrated Infrastructure Initiative in Photonic and Quantum Sciences' – I-PHOQS (IR0000016); the 'Quantum Optical Networks based on Exciton-polaritons' (Q-ONE) funded by the HORIZON-EIC-2022-PATHFINDER CHALLENGES EU programme under grant agreement No. 101115575; the 'Neuromorphic Polariton Accelerator' (PolArt) funded by the Horizon-EIC-2023-Pathfinder Open

EU programme under grant agreement No. 101130304; the PRIN 2022-MUR NanoPix Project (2022YM3232) funded by the European Community-Next Generation EU, Missione 4 Componente 1 (CUP B53D23004650006); and the European Union - NextGeneration EU project "Network 4 Energy Sustainable Transition – NEST" (Project code PE0000021, CUP B53C22004060006, Concession Decree No. 1561 of 11.10.2022 adopted by Ministero dell'Università e della Ricerca).

Views and opinions expressed are however those of the authors only and do not necessarily reflect those of the European Union or European Innovation Council and SMEs Executive Agency (EISMEA). Neither the European Union nor the granting authority can be held responsible for them.

# Supporting Information: Microfluidic-Assisted Growth of Perovskite Microwires for Room-Temperature All-Optical Switching Based on Total Internal Reflection

Annalisa Coriolano[1†], Antonio Fieramosca[1†*], Laura Polimeno[1], Rosanna Mastria[1*], Francesco Todisco[1], Milena De Giorgi[1], Luisa De Marco[1], Aurora Rizzo,[1] Dario Ballarini[1], Ilenia Viola[1,2], Daniele Sanvitto[1]

**Microfluidic-confined growth**

The PDMS replica is transferred onto the ultimate substrate through a dry transfer technique using a home-built micromanipulator. A schematic representation of the setup is depicted in Figure S1. The xy micrometer stage is equipped with a rotator to facilitate the alignment between the PDMS channels and the gold gratings.

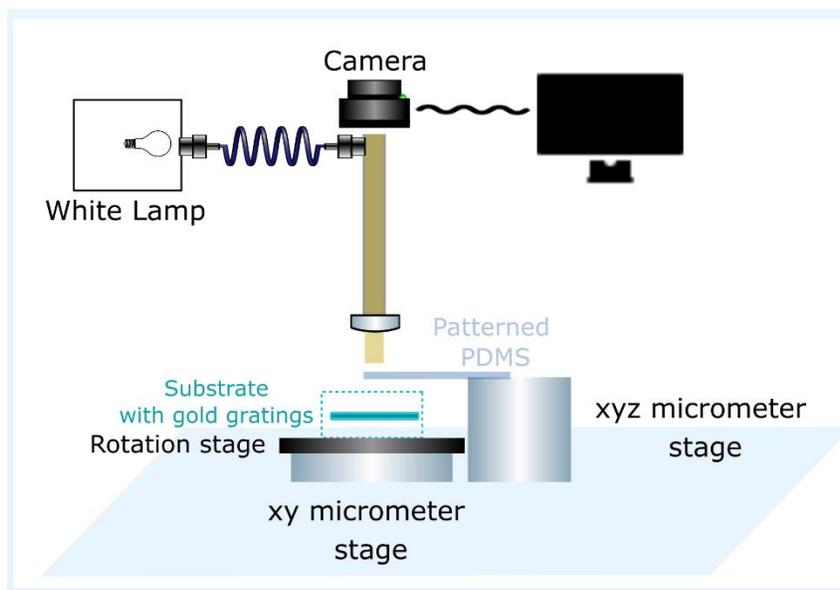

**Figure S1.** Sketch of the home-built micromanipulator used to align the PDMS replica with the gold gratings.

*PDMS template fabrication*

The patterned PDMS template was obtained starting from a silicon master (ThunderNIL) having an array of microchannels with a height (h) of 500 nm and a width (w) ranging from 1 to 20 µm.

PDMS (Sylgard 184, Dow-Corning) was obtained by mixing base and curing agent in a 9:1 ratio and kept it in a desiccator for 30 minutes. After all air bubbles have been carefully removed from PDMS, the liquid elastomer was cast on the silicon master. The system was then cured in an oven for 15 minutes at 140°C. Finally, the elastomeric replica is peeled off from the master thus obtaining a patterned microfluidic device.

*C12 MWs growth*

Figure S2 shows SEM images of the C12 PVK MWs single crystals obtained using the microfluidic-assisted approach for growth on a glass substrate. In particular, we used a template with multiple channels of different lateral dimensions, demonstrating the versatility of the technique in obtaining crystals with desired shapes. The image reveals a flat and uniform surface with sharp edges, free from defects.

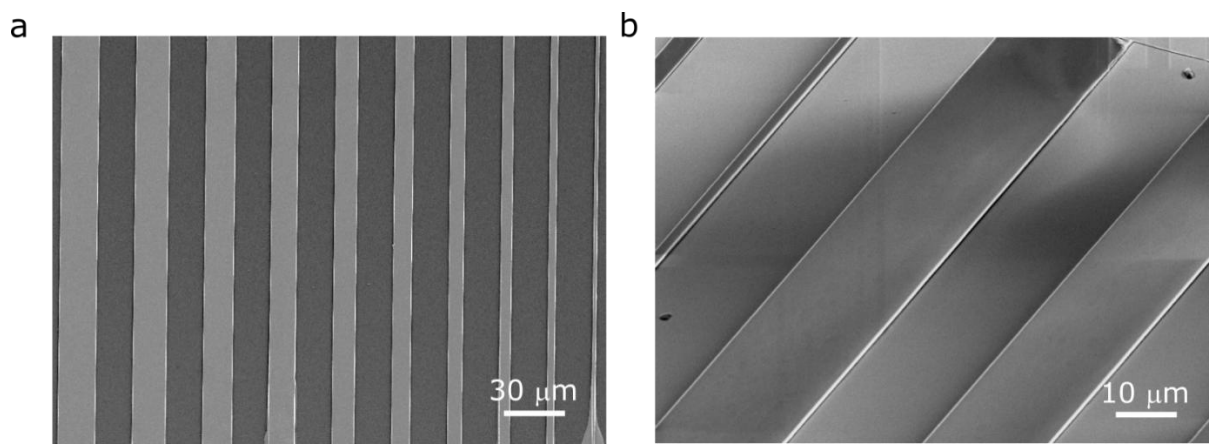

**Figure S2.** SEM images of the C12 PVK MWs on glass substrate.

## Theoretical energy versus in-plane momentum k map

Figure S3 presents the theoretical reflectivity map, for TE polarization, calculated using the same parameters as those used for the map shown in Figure 2 of the main text. This map illustrates the relationship between energy and in-plane momentum ($k$) over a broader energy range, enabling a clearer visualization of the upper polariton branches, otherwise hidden by the high absorption of perovskite in the experimental measurement.

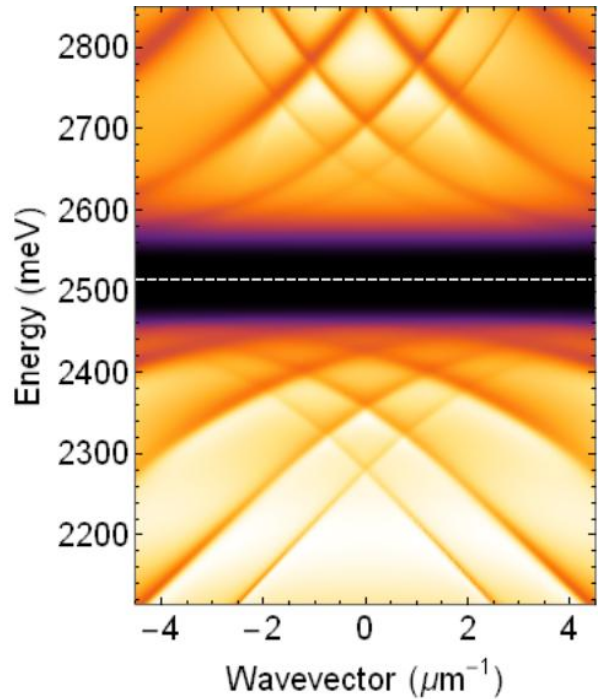

**Figure S3.** Simulated reflectivity map showing Energy versus In-plane momentum ($k$), performed over a broader energy range to reveal the upper polariton branches. The white dashed line indicates the exciton position ($E_{exc}$ = 2525 meV).

## Theoretical reflectivity of the bare the 1-D gold grating

Figure S4 shows the simulated reflectivity map of Energy versus In-plane momentum, $k$, for the bare grating in the same spectral region as the dispersions reported in Figure S3, for TE polarization. This map does not display the formation of any lattice modes, thus confirming that the adopted design effectively functions as an out-coupler for the TIR-confined waveguide polariton modes.

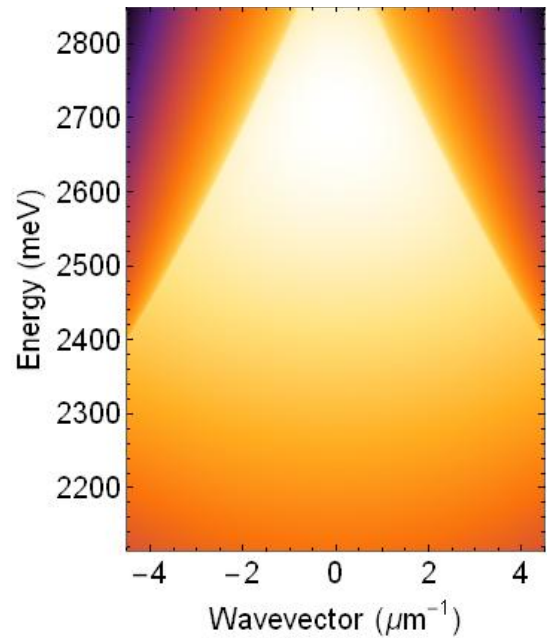

**Figure S4.** Theoretical reflectivity map of Energy versus in-plane momentum $k$ of the bare 1-D gold grating on glass substrate.

**Real space PL maps**

From the real space PL maps, we have estimated the emission intensity collected below the outcoupling region for each distance $d_i$. By doing so, it is possible to evaluate the propagation loss coefficient $\alpha$ in function of the relative distance $\Delta L$, reported in Figure S5, obtaining an average value of $\alpha$ = 0.12 +/- 0.03 dB/um.

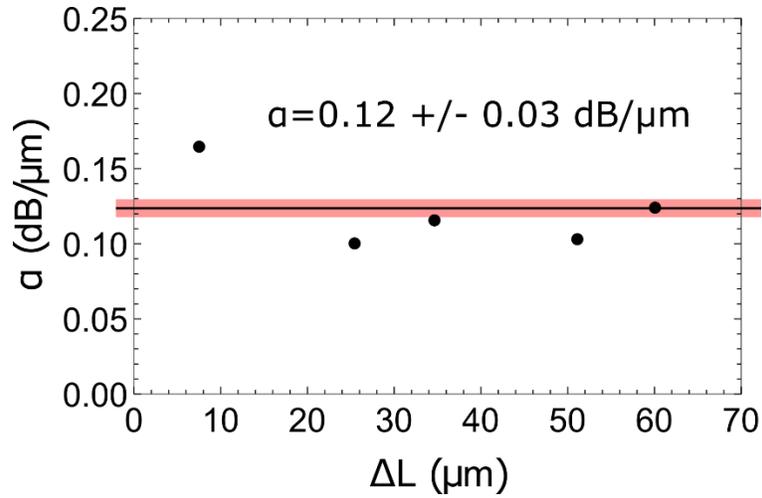

**Figure S5**. Propagation loss ($\alpha$) as a function of the relative distance $\Delta L = (d_i - d_0)$ for each pair of measurements reported in Figure 2d of the main text.

Figure S6 shows a zoom-in of the energy versus in-plane momentum map presented in Figure 2b of the main text. The pumped area is highlighted by a green circle in the zoomed-in energy versus $k_{//}$ map displayed in the figure.

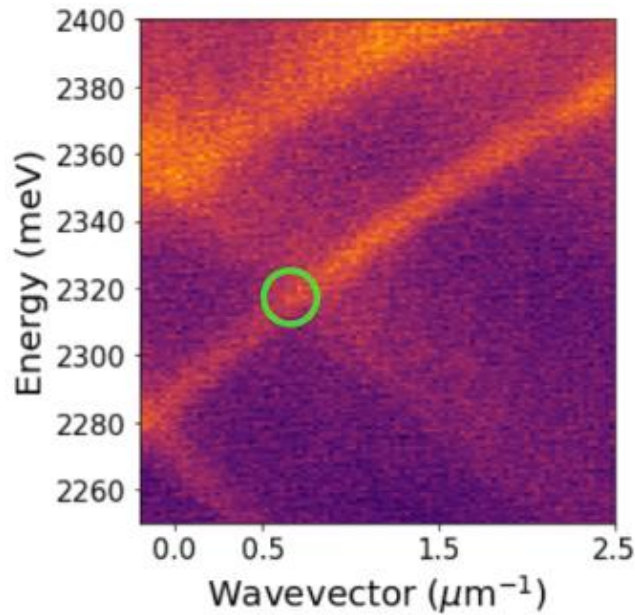

**Figure S6.** Zoom-in of the energy vs in-plane momentum map, with the excited area marked by the green circle.

**Polariton nonlinearities**

The nonlinear response under optical pumping in our samples is investigated by resonantly pumping one of the two gratings with a pulsed laser (10 kHz, 100 fs) in a reflection configuration (Figure S7). Zoomed-in views are shown for low-power excitation (Figure S7b) and high-power excitation (Figure S7c). The LP waveguide modes appear as dips in the laser spectrum, as observed in the corresponding reflection spectra obtained by taking a vertical slice (integrated over 3 pixels) at $k\sim0.7~\mu m^{-1}$, collected at low excitation power (Figure S7d) and high excitation power (Figure S7e). An increase in pumping power leads to a blueshift of the LP branches.

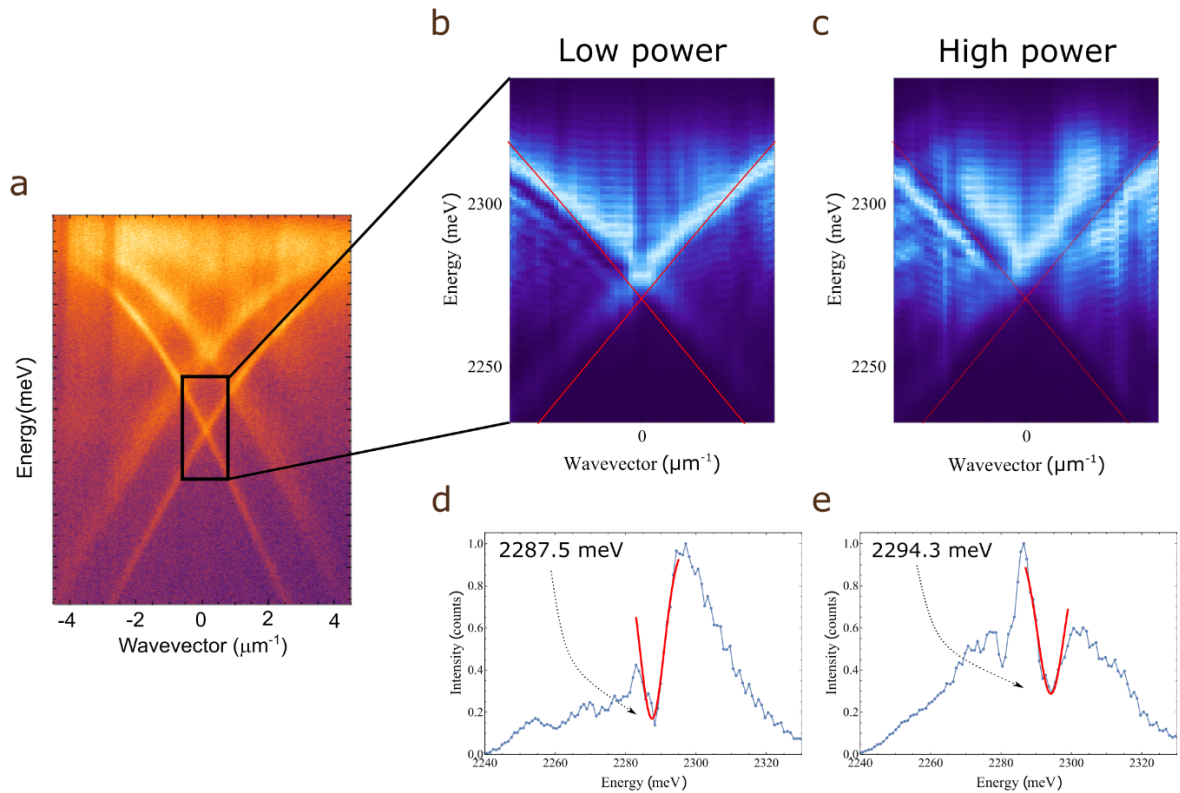

**Figure S7.** (a) Experimental energy versus in-plane momentum photoluminescence map. The white rectangle indicates the pumped area, with a zoomed-in view shown for low-power excitation (b) and high-power excitation (c). The corresponding reflection spectra obtained by taking a vertical slice (integrated over 3 pixels) at $k\sim0.7~\mu m^{-1}$ are displayed for the low-power regime (d) and the high-power regime (e).